\documentclass{article}

\catcode `\@=11
\@addtoreset{equation}{section}

\begin{document}

\def\refname{REFERENCES}

\hsize 140mm
\textheight 190mm

\title{\bf The Corepresentations of Continuous Groups}

\author{\bf J. Koci\'nski $^1$
 and M. Wierzbicki $^2$\\
 \\ \small
$^1$ Grzybowska 5, m.~401, 00-132 Warszawa, Poland\\\small e-mail: \tt kocinsk@if.pw.edu.pl
\\
\\
\small $^2$ Faculty of Physics, Warsaw University of Technology\\
\small Koszykowa 75, 00-772, Warszawa, Poland\\
\small e-mail: \tt wierzba@if.pw.edu.pl
 } 

\date{}

\maketitle

\section{\bf Introduction}

The theory of corepresentations of non-unitary groups 
${\cal G}=G+a_0G$, where $G$ denotes a unitary group and $a_0$ 
is an antiunitary element, was formulated 
by Wigner \cite{Wigner}, to whom belong the first
applications of corepresentations in quantum mechanics. 
Space groups with antiunitary operations and their corepresentations subsequently
found important applications in solid state physics. Lie groups 
with antilinear operations were not considered. 
\\
\indent
Wigner's theory of corepresentations
was elaborated by a number of authors to the form of a powerful tool for
investigations of physical properties of crystals and of magnetic crystals \cite{Birman, Bir,
Bradley, Dimmock, Kovalev6, Kovalev7}.
It was applied in the investigations of symmetry changes at 
commensurate and incommensurate continuous magnetic
phase transitions \cite{Jar1, Kocinski1, Kocinski2, Kocinski3}, and to the problem of
magnetocrystalline anisotropy of ferromagnetic crystals \cite {KocinskiP}.
\\
\indent
It was shown by Birman \cite{Birman} that a non-unitary symmetry group can intervene 
in the classical description of a crystal in a state of thermodynamic equilibrium. 
The non-unitary group of the type $G+KG$,  where $G$ is a space group, and 
where $K$ denotes the operation of complex conjugation, 
constitutes the complete symmetry group of the  crystal-lattice dynamic problem. 
This group plays the basic role in establishing the
one-to-one correspondence between vibration frequencies and irreducible corepresentations
(coirreps). 
The application of the non-unitary group $G+KG$ to a description of lattice vibrations
elaborated in \cite{Birman}, opened the way for a
further development in this field, made by Kovalev \cite{Kovalev1, Kovalev2, Kovalev3,
Kovalev4, Kovalev7} and by
Kovalev and Gorbanyuk \cite{Kovalev6}. These authors formulated another method 
of demonstrating that there 
exists the one-to-one correspondence between coirreps and frequencies of crystal 
lattice vibrations. As a subsequent step in the exploration of the 
importance of the non-unitary groups, Kovalev and Gorbanyuk \cite{Kovalev6},
generalized Wigner-Eckart theorem \cite{Streitwolf} to systems described by magnetic
space groups (see \cite{Kocinski3}).
\\
\indent
The corepresentation theory was originally formulated for the case when the subgroup
$G$ of the group $G+a_0G$ is unitary \cite{Wigner}. 
The group $G+a_0G$ then is called a non-unitary group \cite{Bradley}, 
and the element $a_0$ is an antiunitary operation. 
The name antiunitary, which was assigned to 
the antilinear operations of complex conjugation $K$ and of time reversal $\Theta$ 
draws from the fact that when the bilinear product of basis functions is Hermitian, any
antiunitary operation is equal to the product of the operation of complex conjugation $K$
with some unitary operation \cite{Wigner}.
The name antiunitary does not seem to be appropriate when the operations $K$ or $\Theta$ 
are applied to linear operations which are not unitary,
for example to the operations of the group $SL(2,C)$.  The modification of group
representation theory leading to corepresentations is conditioned by the antilinear character
of the operations $K$ or $\Theta$.
\\
\indent
In Section 2 we will present the theory of corepresentations without making the assumption
that the subgroup $G$ of the group $G+a_0G$ is unitary.
Groups of the type ${\cal G}=G+a_0G$ will be considered,
consisting of the subgroup $G$ which is a group of linear operations and of
the coset $a_0G$, consisting of products of an antilinear operation $a_0$ with the
linear operations
belonging to $G$. The element $a_0$ itself, in general can be a product of an antilinear 
operation $A$ with a linear operation $g^0_L$,
which does not belong to the subgroup $G$. However, the element $g^0_L$ has to be of
such a type that we have $(Ag^0_L)^2\in G$. In particular we can have $g^0_L$ equal to 
the unit element ${\bf 1}$.

\section {\bf The corepresentation theory of continuous groups}

In this Section we are indebted to the presentations of 
the corepresentation theory for magnetic space groups by Bradley and Cracknell \cite{Bradley},
and Kovalev and Gorbanyuk \cite{Kovalev6}. 
\\
\indent
In the applications of corepresentation theory in quantum mechanics \cite{Wigner},
the antiunitary
element $a_0$ was the operation of time-reversal
$\Theta$,
multiplied by a proper or improper rotation element, represented by a unitary matrix. 
When the subgroup $G$ need not be unitary, it seems to be misleading to call the 
group $G+a_0G$ a non-unitary group, and we will not use this name.
\\
\indent
Let $G$ be a continuous group of linear transformations which need not be unitary. 
We define the group
\begin{equation}
{\cal G}=G+a_0G
\label{eq2:1}
\end{equation}

\noindent
in which in general the operation $a_0$  is a product of an antilinear operation with a linear
operation, which does not belong to the subgroup $G$. 
As the product of any two elements of the coset $a_0G$ has to belong to $G$, 
we must have $a^2_0\in G$.
\\
\indent
Let $\Gamma$ be an irreducible representation (irrep) of the group $G$, of dimension $d$, and 
let $\varphi_i,\,i=1,...,d$, be its basis functions. For any element $g\in G$, we then
have

\begin{equation}
g\varphi_i=\sum\limits_{j=1}^d\Delta(g)_{ji}\varphi_j
\quad\mbox{or}\quad
g\varphi=\tilde\Delta(g)\varphi
\label{eq2:2}
\end{equation}

\noindent
where $\Delta(g)$ is the representation matrix, $\varphi$ is the column matrix
constructed from the basis functions $\varphi_1,\,...,\,\varphi_d$, and $\tilde\Delta(g)$ is the transposed matrix.
The action of an antilinear operation $a_0$ on a linear combination of functions $\varphi_i$ is defined by

\begin{equation}
a_0\sum_{i=1}^d c_i\varphi_i=\sum_{i=1}^d c_i^* a_0\varphi_i
\label{eq2:3}
\end{equation}

\noindent
where $c_i$ are complex numbers, and $*$ denotes complex conjugation.
\\
\indent
The action of the antilinear element $a_0$ on the basis functions $\varphi_i$ leads to
another set of functions $\phi_i$,
\begin{equation}
a_0\varphi_i=\phi_i; \quad i=1,...,d
\label{eq2:4}
\end{equation}

\noindent
We consider this transformation as an endomorphism of the space which is spanned by the functions $\varphi_i$.
The column matrix constructed from the functions $\phi_i,\,i=1,2,...,d$
will be denoted by $\phi$. The action of $g\in G$ on $\phi$ is given by
\begin{equation}
g\phi=ga_0\varphi=a_0(a^{-1}_0ga_0)\varphi=a_0g^{\prime}\varphi=a_0\tilde\Delta(g^{\prime})
\varphi=a_0\tilde\Delta(a^{-1}_0ga_0)\varphi=\tilde\Delta^{\ast}(a^{-1}_0ga_0)\phi
\label{eq2:5}
\end{equation}

\noindent
where the last equality is connected with the antilinear character of $a_0$. From Eqs. 
(\ref{eq2:2}) and (\ref{eq2:5})
we obtain

\begin{equation}
g\left(\begin{array}{r}
\varphi
\\
\phi
\end{array}\right)=\left(\begin{array}{r|r}
\tilde\Delta(g) & 0
\\
\hline
0 & \tilde\Delta^{\ast}(a^{-1}_0ga_0)
\end{array}\right)\left(\begin{array}{r}
\varphi
\\
\phi
\end{array}\right); \quad \forall g\in G
\label{eq2:6}
\end{equation}

\noindent
We now define the matrix $\overline\Delta(g)$ in the representation $\overline\Gamma$, by

\begin{equation}
\overline\Delta(g)=\Delta^{\ast}(a^{-1}_0ga_0); \quad \overline\Delta(g)\in\overline\Gamma
\label{eq2:7}
\end{equation}

\noindent
where $\overline\Delta(g)$ is a matrix representative of $g\in G$, in the representation
$\overline\Gamma$ of $G$. This equation defines the representation $\overline\Gamma$ of $G$.
\\
\indent
Let $a$ be any element of $a_0G$, say, $a_0g$. We then obtain
\begin{equation}
a\varphi=a_0g\varphi=a_0\tilde\Delta(g)\varphi=\tilde\Delta^{\ast}(g)\phi=\tilde\Delta^{\ast}
(a^{-1}_0a)\phi
\label{eq2:8}
\end{equation}

\noindent
where Eqs. (\ref{eq2:2}) and (\ref{eq2:4}) and the antilinear character of $a_0$ have been used.
We next obtain
\begin{equation}
a\phi=aa_0\varphi=\tilde\Delta(aa_0)\varphi
\label{eq2:9}
\end{equation} 

\noindent
owing to $aa_0\in G$. From Eqs. (\ref{eq2:8}) and (\ref{eq2:9}) we obtain
the expression

\begin{equation}
a\left(\begin{array}{r}
\varphi
\\
\phi
\end{array}\right)=\left(\begin{array}{r|r}
0 & \tilde\Delta^{\ast}(a^{-1}_0a)
\\
\hline
\tilde\Delta(aa_0) & 0
\end{array}\right)\left(\begin{array}{r}
\varphi
\\
\phi
\end{array}\right)
\label{eq2:10}
\end{equation}

\noindent
If $a=ga_0$, and $g=aa^{-1}_0$, the same formula is obtained, since we have
$$
a\varphi=(ga_0)\varphi=\tilde\Delta^{\ast}(a^{-1}_0ga_0)\phi=\tilde\Delta^{\ast}(a^{-1}_0a)\phi
$$
\noindent
and
\noindent
$$
a\phi=aa_0\varphi=\tilde\Delta(aa_0)\varphi
$$

\noindent
which are Eqs. (\ref{eq2:8}) and (\ref{eq2:9}), respectively. Equations (\ref{eq2:6}) 
and (\ref{eq2:10}) demonstrate the invariance of 
the space spanned by the functions $\varphi_i$ and $\phi_i$, $\,i=1,...,d$, under the group
${\cal G}$. From Eqs. (\ref{eq2:6}) and (\ref{eq2:10}) we obtain the matrices

\begin{equation}
D(g)=\left(\begin{array}{r|r}
\Delta(g) & 0
\\
\hline
0 & \Delta^{\ast}(a^{-1}_0ga_0)
\end{array}\right);\quad\forall g\in G
\label{eq2:11}
\end{equation}

\noindent
and

\begin{equation}
D(a)=\left(\begin{array}{r|r}
0 & \Delta(aa_0)
\\
\hline
\Delta^{\ast}(a^{-1}_0a) & 0
\end{array}\right);\quad a=a_0g\quad\mbox{or}\quad a=ga_0,\quad\forall g\in G
\label{eq2:12}
\end{equation}

\noindent
The sets of matrices in Eqs. (\ref{eq2:11}) and (\ref{eq2:12}) form the {\em corepresentation 
of the group ${\cal G}$},
derived from the representation $\Gamma$, with the matrices $\Delta(g)$ of the subgroup $G$.
This corepresentation may be reducible.
The corepresentation matrices obey the following set of equations \cite{Wigner}:
\begin{equation}
\begin{array}{l}
D(g_1)D(g_2)=D(g_1g_2);\quad\forall g_1,g_2\in G\\

D(g)D(a)=D(ga);\quad\forall g\in G,\,\mbox{and}\,\,\forall a\in a_0G\\

D(a)D^{\ast}(g)=D(ag);\quad\forall g\in G,\,\mbox{and}\,\,\forall a\in a_0G\\

D(a_1)D^{\ast}(a_2)=D(a_1a_2);\quad\forall a_1,a_2\in a_0G
\end{array}
\label{eq2:13}
\end{equation}

\noindent
These are established by examining the action of the respective products of elements $g$ and $a$ 
on the basis functions, when the antilinear character of the elements $a$ is taken into account.
Because of the last two equalities, the mapping ${\cal G}\rightarrow
D\Gamma$ is not a homomorphism. 
\\
\\
\noindent
{\bf The equivalence of two corepresentations}.
\\
\noindent
Performing the basis transformation with a nonsingular transformation $S$,
\begin{equation}
\tilde S\chi=\chi^{\prime},\quad\mbox{with}\quad\tilde\chi=(\varphi, \phi)
\label{eq2:14}
\end{equation}

\noindent
where $\varphi$ and $\phi$ are given in Eq. (\ref{eq2:6}), we obtain

\begin{equation}
g\chi^{\prime}=\tilde D^{\prime}(g)\chi^{\prime},\quad {\rm hence}\quad
D^{\prime}(g)=S^{-1}D(g)S
\label{eq2:15}
\end{equation}

\noindent
and 
\begin{equation}
a\chi^{\prime}=\tilde D^{\prime}(a)\tilde S\chi,\quad
{\rm or}\quad a\chi^{\prime}=a\tilde S\chi=\tilde S^{\ast}\tilde D(a)\chi,\quad {\rm hence}
\quad D^{\prime}(a)=S^{-1}D(a)S^{\ast}
\label{eq2:16}
\end{equation}

\noindent
Two corepresentations of the group ${\cal G}$, the corepresentation with the matrices $D(g)$ and $D(a)$, and the corepresentation with the matrices $D'(g)$ and
$D'(a)$
are said to be equivalent if there exists a nonsingular matrix $S$ such that

\begin{equation}
D^{\prime}(g)=S^{-1}D(g)S;\qquad \forall g\in G
\label{eq2:17}
\end{equation}
\begin{equation}
D^{\prime}(a)=S^{-1}D(a)S^{\ast};\qquad \forall a\in a_0G
\label{eq2:18}
\end{equation}

\noindent
The equivalence conditions in Eqs. (\ref{eq2:17}) and (\ref{eq2:18}) allow to replace 
the corepresentation matrices $D(a)$, by the matrices ${\rm exp}(i\alpha_0)D(a)$, with a 
real parameter $\alpha_0$, by applying the transformation
$S={\rm exp}(-i\alpha_0/2)E$, where $E$ denotes the unit matrix. 
The matrices $D(g)$ remain unaltered. 
The matrices of the elements $a$ then acquire the form
\begin{equation}
D^{\prime}(a)={\rm e}^{i\alpha_0}D(a)
\label{eq2:19}
\end{equation}

\noindent
It can be shown that
there is no ambiguity in the assignment of the corepresentation $D\Gamma$,
derived from the representation $\Gamma$, to the group ${\cal G}$. Different choices of $a_0$
in the definition of ${\cal G}$ lead to equivalent corepresentations \cite{Bradley, Wigner}.
\\
\\
\noindent
{\bf Reducibility of corepresentations.}
\\
\noindent
If the basis $\chi$ in Eq. (\ref{eq2:14}) can be transformed by a nonsingular transformation 
$S$ so that the 
new basis $\chi^{\prime}=\tilde S\chi$ is the direct sum of two subspaces which are both
invariant under the group ${\cal G}$, the corep $D\Gamma$ is said to be {\em reducible}.
If not, $D\Gamma$ is said to be {\em irreducible}. In the case of irreducibility, 
all the matrices 
of the corep $D^{\prime}\Gamma$, equivalent to $D\Gamma$, must be in the same block-diagonal
form.
The two representations $\Gamma$ and $\overline\Gamma$ may be inequivalent or equivalent.
The answer to the question about
the reducibility of the corep in Eqs. (\ref{eq2:11}) and (\ref{eq2:12}) hinges upon that.
\\
\\
\noindent
{\bf The representations $\bf\Gamma$ and $\bf\overline\Gamma$ are inequivalent}.
\\
\noindent
Let us suppose that the corep matrices in Eqs. (\ref{eq2:11}) and (\ref{eq2:12}) are reduced 
by a matrix $S$. Since the matrices $D(g),\,g\in G$, are in the form of the direct sum of 
the irreducible matrices $\Delta(g)$ and $\Delta^{\ast}(a^{-1}_0ga_0)$, their only 
reduced form is
$$
\left(\begin{array}{r|r}
X(g) & 0
\\
\hline
0 & Y(g)
\end{array}\right)
$$

\noindent
where the matrices $X(g)$ and $Y(g)$ are equivalent to $\Delta(g)$ and $\Delta^{\ast}
(a^{-1}_0ga_0)$, respectively. Writing
$$
S^{-1}=\left(\begin{array}{rr}
p & q
\\
r & s
\end{array}\right)
$$

\noindent
we obtain the condition in Eq. (\ref{eq2:17}) in the form
$$
\left(\begin{array}{rr}
p & q
\\
r & s
\end{array}\right)\left(\begin{array}{r|r}\Delta(g) & 0
\\
\hline
0 & \Delta^{\ast}(a^{-1}_0ga_0)
\end{array}\right)=\left(\begin{array}{r|r}
X(g) & 0
\\
\hline
0 & Y(g)
\end{array}\right)\left(\begin{array}{rr}
p & q
\\
r & s
\end{array}\right)
$$

\noindent
from which we obtain the three conditions,
$$
(1)\quad p\Delta(g)=X(g)p;\quad (2)\quad s\Delta^{\ast}(a^{-1}_0ga_0)=Y(g)s
$$
and 
$$
(3)\quad q\Delta^{\ast}(a^{-1}_0ga_0)=X(g)q=p\Delta(g)p^{-1}q
$$

\noindent
where in the last equality we made use of the fact that $p^{-1}$ must exist, as it provides
the equivalence transformation between $\Delta(g)$ and $X(g)$. From (1) and (3) we find that
$$
(p^{-1}q)\Delta^{\ast}(a^{-1}_0ga_0)=\Delta(g)(p^{-1}q)
$$

\noindent
As $\Delta(g)$ and 
$\Delta^{\ast}(a^{-1}_0ga_0)$ were assumed to be inequivalent, by Schur's Lemma we must
have $p^{-1}q=0$, and, consequently, $q=0$. In an analogous way we can find that $r=0$, and then
$S^{-1}$ comes out in the form
$$
S^{-1}=\left(\begin{array}{rr}
p & 0
\\
0 & s
\end{array}\right)
$$ 

\noindent
This matrix cannot reduce the matrices $D(a)$ in Eq. (\ref{eq2:12}) to a block-diagonal form.
Consequently,
if the irreps $\Gamma$ and $\overline\Gamma$ are inequivalent, the corep of the group
${\cal G}$ derived from the irrep $\Gamma$ is irreducible. We are dealing with $c-$type
irreducible corepresentation (type 3 in \cite{Wigner}), with the matrices in 
Eqs. (\ref{eq2:11}) and (\ref{eq2:12}.)
\\
\\
\noindent
{\bf The representations $\bf\Gamma$ and $\bf\overline\Gamma$ are equivalent}.
\\
\noindent
There exists then a nonsingular matrix $N$ (the matrix
$\bf\beta$ in \cite{Wigner}) such that 

\begin{equation}
\Delta(g)=N\Delta^{\ast}(a^{-1}_0ga_0)N^{-1},\quad\forall g\in G
\label{eq2:20}
\end{equation}

\noindent
Replacing the element $g$ with $a^{-1}_0ga_0$ we also obtain 
\begin{equation}
\Delta^{\ast}(a^{-1}_0ga_0)=N^{\ast}\Delta(a^{-2}_0ga^2_0)(N^{-1})^{\ast}=
N^{\ast}\Delta(a^{-2}_0)\Delta(g)\Delta(a^2_0)(N^{-1})^{\ast}
\label{eq2:21}
\end{equation}

\noindent
Substituting the last expression into Eq. (\ref{eq2:20}) we obtain the equation

\begin{equation}
\Delta(g)=NN^{\ast}\Delta^{-1}(a^2_0)\Delta(g)\Delta(a^2_0)(N^{-1})^{\ast}N^{-1},\quad
\forall g\in G
\label{eq2:22}
\end{equation}

\noindent
Since $\Gamma$ is irreducible, it follows from Schur's Lemma that
$NN^{\ast}\Delta^{-1}(a^2_0)=\Lambda E$
where $\Lambda$ is a constant and $E$ is the unit matrix. Hence we obtain:
\begin{equation}
\Delta(a^2_0)=\Lambda^{-1}NN^{\ast},\quad {\rm and}\quad \Delta^{\ast}(a^2_0)=
(\Lambda^{\ast})^{-1}N^{\ast}N
\label{eq2:23}
\end{equation}

\noindent
In Eq. (\ref{eq2:20}) we can put $g=a^2_0$ and we then obtain

\begin{equation}
\Delta(a^2_0)=N\Delta^{\ast}(a^2_0)N^{-1}
\label{eq2:24}
\end{equation}

\noindent
Substituting the right hand sides of Eqs. (\ref{eq2:23}) into 
Eq. (\ref{eq2:24}), we obtain the equalities: 
$\Lambda^{-1}NN^{\ast}=N(\Lambda^{\ast})^{-1}N^{\ast}NN^{-1}=(\Lambda^{\ast})^{-1}NN^{\ast}$,
and hence $\Lambda=\Lambda^{\ast}$. Calculating the determinant of both sides of
Eqs. (\ref{eq2:23}) we obtain:

\begin{equation}
\Lambda=\pm \frac{|{\rm det}N{\rm det}N^{\ast}|}{|{\rm det}\Delta(a^2_0)|}=\pm 1
\label{eq2:25}
\end{equation}

\noindent
when we assume that the irrep $\Gamma$ consists of matrices with $|{\rm det}\Delta (g)=1|$,
and we remember that the matrix $N$ can always be chosen so as to have
$|{\rm det}N{\rm det}N^{\ast}|=1$. Consequently, from Eq. (\ref{eq2:23}) we obtain

\begin{equation}
NN^{\ast}=\pm\Delta(a^2_0)
\label{eq2:26}
\end{equation}

\noindent
as in the case of unitary matrices $N$, as in \cite{Wigner},\cite{Bradley},\cite{Kocinski3}.
The reducibility of a corep depends on the sign in Eq. (\ref{eq2:26}).
\\
\indent
A corepresentation $D\Gamma$ is reducible if and only if the matrices $D(g)$ and $D(a)$ 
can simultaneously be expressed  in the same block-diagonal form. The matrices 
$D(g)$ in Eq. (\ref{eq2:11}) are already
in a reduced form, however, it will be convenient to convert them to the 
form, when there are the same  blocks along the diagonal. 
Applying the matrix

\begin{equation}
W=\left(\begin{array}{r|r}
E & 0
\\
\hline
0 & -N^{-1}
\end{array}\right)
\label{eq2:27}
\end{equation}

\noindent
with $N$ from Eq. (\ref{eq2:20}), and $D(a_0)$ in Eq. (\ref{eq2:12}), we obtain:

\begin{equation}
D^{\prime}(g)=W^{-1}D(g)W=
\left(\begin{array}{r|r}
\Delta(g) & 0
\\
\hline
0 & \Delta(g)
\end{array}\right)
\label{eq2:28}
\end{equation}

\noindent
and

\begin{equation}
D^{\prime}(a_0)=W^{-1}D(a_0)W^{\ast}=
\left(\begin{array}{r|r}
0 & -\Delta(a^2_0)(N^{-1})^{\ast}
\\
\hline
-N & 0
\end{array}\right)
\label{eq2:29}
\end{equation}

\noindent
Since every element of ${\cal G}$ is of the form $g, a_0g$ or $ga_0$, for $g\in G$, while
$D^{\prime}(a_0g)=D^{\prime}(a_0)D^{\prime\,\ast}(g)$ and $D^{\prime}(ga_0)=
D^{\prime}(g)D^{\prime}(a_0)$, a nonsingular transformation $V$ is required, which will reduce 
the matrices $D^{\prime}(a_0)$ to block-diagonal form,
leaving the matrices $D^{\prime}(g)$ unaltered. That $V$ must commute with
$D^{\prime}(g)$ in Eq. (\ref{eq2:28}). Writing:

\begin{equation}
V^{-1}=\left(\begin{array}{rr}
\alpha & \beta
\\
\gamma & \delta
\end{array}\right)
\label{eq2:30}
\end{equation}

\noindent
from the equation $V^{-1}D^{\prime}(g)=D^{\prime}(g)V^{-1}$ we obtain:

\begin{equation}
\left(\begin{array}{rr}
\alpha\Delta(g) & \beta\Delta(g)
\\
\gamma\Delta(g) & \delta\Delta(g)
\end{array}\right)=\left(\begin{array}{rr}
\Delta(g)\alpha & \Delta(g)\beta
\\
\Delta(g)\gamma & \Delta(g)\delta
\end{array}\right)
\label{eq2:31}
\end{equation}

\noindent
As the matrices $\Delta(g)$ are irreducible, from Schur's Lemma we find that
$\alpha=\lambda E, \,\beta=\mu E,\, \gamma=\nu E$ and $\delta=\rho E$, with constant
$\lambda,\mu,\nu,\rho$, where $E$ is a $d-$dimensional unit matrix. We therefore must have

\begin{equation}
V^{-1}=\left(\begin{array}{rr}
\lambda E & \mu E
\\
\nu E & \rho E
\end{array}\right)
\label{eq2:32}
\end{equation}

\noindent
The required existence of $V$ implies that ${\rm det}\, V^{-1}\neq 0$, which leads to

\begin{equation}
\lambda\rho\neq\mu\nu
\label{eq2:33}
\end{equation}

\noindent
We find that

\begin{equation}
V=\frac{1}{2}\left(\begin{array}{rr}
E/\lambda & E/\nu
\\
E/\mu & E/\rho
\end{array}\right)=E
\label{eq2:34}
\end{equation}

\noindent
where $E$ in the matrix is a $d-$dimensional unit matrix, and on the right hand side $E$ 
is a $2d-$dimensional unit matrix, with

\begin{equation}
\lambda\rho=-\mu\nu
\label{eq2:35}
\end{equation}

\noindent
which is the condition for a reduction of a corep to be possible. With $D^{\prime}(a_0)$
in Eq. (\ref{eq2:29}), the transformed matrix $D^{\prime\prime}(a_0)$ has the form

\begin{eqnarray}
D^{\prime\prime}(a_0)=V^{-1}D^{\prime}(a_0)V^{\ast}=
\nonumber\\
\frac{1}{2}\left(\begin{array}{r|r}
-(\mu/\lambda^{\ast})N-(\lambda/\mu^{\ast})\Delta(a^2_0)(N^{-1})^{\ast} & 
-(\mu/\nu^{\ast})N-(\lambda/\rho^{\ast})\Delta(a^2_0)(N^{-1})^{\ast}
\\
\hline
-(\rho/\lambda^{\ast})N-(\nu/\mu^{\ast})\Delta(a^2_0)(N^{-1})^{\ast} &
-(\rho/\nu^{\ast})N-(\nu/\rho^{\ast})\Delta(a^2_0)(N^{-1})^{\ast}
\end{array}\right)
\label{eq2:36}
\end{eqnarray}

\noindent
As the off-diagonal terms have to vanish, and from
Eq. (\ref{eq2:35}), we obtain the condition:

\begin{equation}
NN^{\ast}=\frac{|\lambda|^2}{|\mu|^2}\Delta(a^2_0)
\label{eq2:37}
\end{equation}

\noindent
which has the form of Eq.(\ref{eq2:26}) with $(+)$ sign, provided that

\begin{equation}
\frac{|\lambda|^2}{|\mu|^2}=1
\label{eq2:38}
\end{equation}

\noindent
{\bf The irreps $\Gamma$ and $\overline\Gamma$ are equivalent and
a reduction of the corepresentation in Eqs. (\ref{eq2:11}) and
(\ref{eq2:12}) is possible}.
\\
\noindent
Considering Eqs. (\ref{eq2:37}) and (\ref{eq2:38}) we find that
a reduction of the corep in Eqs. (\ref{eq2:11}) and (\ref{eq2:12}) is possible when 

\begin{equation}
NN^{\ast}=+\Delta(a^2_0)
\label{eq2:39}
\end{equation}

\noindent
Taking into account Eqs. (\ref{eq2:35}) and (\ref{eq2:37}), we obtain from Eq. 
(\ref{eq2:36}) the reduced matrix $D^{\prime\prime}(a_0)$ in the form

\begin{equation}
D^{\prime\prime}(a_0)=\left(\begin{array}{r|r}
(-\mu/\lambda^{\ast})N & 0
\\
\hline
0 & (-\rho/\nu^{\ast})N
\end{array}\right)
\label{eq2:40}
\end{equation}

\noindent
Owing to Eq. (\ref{eq2:35}), the coefficients
$\mu/\lambda^{\ast}$ and $\rho/\nu^{\ast}$, have the same absolute
value and they can differ only by a phase factor, 
and hence, according to Eq. (\ref{eq2:18}), the two blocks along the diagonal 
are equivalent. For a
unitary $N$, the matrix in Eq. (\ref{eq2:40}) 
turns into the customary
matrix $D^{\prime\prime}(a_0)$, for example in Eq. (7.3.40) of \cite{Bradley},
or in Eq. (1.5.40) of \cite{Kocinski3}.
\\
\indent
In order to determine the matrix connected with the element $a=ga_0$ we use the second from
Eqs. (\ref{eq2:13}) and obtain
$D(a)=D(ga_0)=D(g)D(a_0)$, hence from Eqs. (\ref{eq2:28}) and (\ref{eq2:40}) 
we obtain the matrix

\begin{equation}
D^{\prime\prime}(ga_0)=\left(\begin{array}{r|r}
-(\mu/\lambda^{\ast})\Delta(g)N & 0
\\
\hline
0 & -(\rho/\nu^{\ast})\Delta(g)N
\end{array}\right)
\label{eq2:41}
\end{equation}

\noindent
We observe that the reduced matrices in Eqs. (\ref{eq2:28}) and (\ref{eq2:41}), with 
the two blocks in $D^{\prime\prime}(a)$ in the same
form, can be obtained by applying to corep matrices in Eqs. (\ref{eq2:11}) and (\ref{eq2:12}) 
the transformation, which is analogous to that given by Kovalev and 
Gorbanyuk for unitary groups \cite{Kovalev6}, namely:

\begin{equation}
V_1=\frac{1}{\sqrt{2}}\left(\begin{array}{r|r}
E & iE
\\
\hline
(\lambda/\mu)N^{-1} & -i(\lambda/\mu)N^{-1}
\end{array}\right), \qquad V^{-1}_1=\frac{1}{\sqrt{2}}\left(\begin{array}{r|r}
E & (\mu/\lambda)N
\\
\hline
-iE & i(\mu/\lambda)N
\end{array}\right)
\label{eq2:42}
\end{equation}

\noindent
Applying this transformation and
taking into account the similarity transformation in Eqs. (\ref{eq2:17}) and  (\ref{eq2:18}), 
we obtain corep matrices in the form:

\begin{equation}
D^{\prime}(g)=\left(\begin{array}{r|r}
\Delta(g) & 0
\\
\hline
0 & \Delta(g)
\end{array}\right),\quad D^{\prime}(ga_0)={\rm e}^{i\alpha_0}\left(\begin{array}{r|r}
(\mu/\lambda)\Delta(g)N & 0
\\
\hline
0 & (\mu/\lambda)\Delta(g)N
\end{array}\right)
\label{eq2:43}
\end{equation}

\noindent
and 

\begin{equation}
D^{\prime}(a_0g)={\rm e}^{i\alpha_0}\left(\begin{array}{r|r}
(\mu/\lambda) N\Delta^{\ast}(g) & 0
\\
\hline
0 & (\mu/\lambda)N\Delta^{\ast}(g)
\end{array}\right)
\label{eq2:44}
\end{equation}

\noindent
With $g=E$, we obtain

\begin{equation}
D^{\prime}(a_0)={\rm e}^{i\alpha_0}\left(\begin{array}{r|r}
(\mu/\lambda)N & 0
\\
\hline
0 & (\mu/\lambda)N
\end{array}\right)
\label{eq2:45}
\end{equation}

\noindent
which replaces Eq. (\ref{eq2:40}), in which the two block matrices appear with opposite signs.
When the matrix $N$ is unitary, and we put $\mu/\lambda=1$, the
transformation $V_1$ in Eq. (\ref{eq2:42}) turns into Eq. (8.11a) in
\cite{Kovalev6}, or into Eq. (1.5.43) in \cite{Kocinski3}.  
In general we have $\mu/\lambda={\rm exp}(i\xi)$, with a real $\xi$, and
this exponential factor can be absorbed by the factor ${\rm exp}(i\alpha_0)$. 
\\
\indent
Renaming the functions $\phi_i$ in Eq. (\ref{eq2:4}) of the original corepresentation,

\begin{equation}
\phi_i=a_0\varphi_i=\varphi_{d+i},\quad i=1,...,d
\label{eq2:46}
\end{equation}

\noindent
and utilizing the transformation $V_1$ in Eq. (\ref{eq2:42}), we obtain the basis functions of
the two blocks, with labels (1) and (2), 

\begin{equation}
\psi^{(1)}_j=\frac{1}{\sqrt{2}}\Big(\varphi_j+\frac{\lambda}{\mu}\sum\limits_{i=1}^d
(\tilde N^{-1})_{ij}
\varphi_{d+i}\Big),\qquad \psi^{(2)}_j=\frac{i}{\sqrt{2}}\Big(\varphi_j-\frac{\lambda}{\mu}
\sum\limits_{i=1}^d (\tilde N^{-1})_{ij}\varphi_{d+i}\Big)
\label{eq2:47}
\end{equation}

\noindent
In the case of unitary groups $G$, when the original basis functions $\varphi_j$ 
in Eq. (\ref{eq2:2}) are orthogonal,
the basis functions $\psi^{(1)}_j ,j=1,...,d$, also are orthogonal, and the same holds for the 
functions $\psi^{(2)}_j$. These two sets of functions need not be mutually orthogonal,
however. 
\\
\\
\noindent
{\bf The irreps $\Gamma$ and $\overline\Gamma$ are equivalent, however a reduction of the 
corepresentation in Eqs. (\ref{eq2:11}) and (\ref{eq2:12}) is impossible}.
\\
\\
\noindent
According to Eq. (\ref{eq2:26}) with the $(-)$ sign, we now have:

\begin{equation}
NN^{\ast} =-\Delta(a^2_0)
\label{eq2:48}
\end{equation}

\noindent
and from Eq. (\ref{eq2:29}) we obtain

\begin{equation}
D^{\prime}(a_0)=\left(\begin{array}{r|r}
0 &  N
\\
\hline
-N & 0
\end{array}\right)
\label{eq2:49}
\end{equation}

\noindent
With $a=ga_0$, hence $g=aa^{-1}_0$, and $D^{\prime}(a)=D^{\prime}(g)D^{\prime}(a_0)$, 
with $D^{\prime}(g)$ in Eq. (\ref{eq2:28}) and $D^{\prime}(a_0)$ in Eq. (\ref{eq2:49}),
we obtain

\begin{equation}
D^{\prime}(ga_0)=\left(\begin{array}{r|r}
0 & \Delta(g)N
\\
\hline
-\Delta(g)N & 0
\end{array}\right)
\label{eq2:50}
\end{equation}

\noindent
With $a=a_0g$, hence $g=a^{-1}_0a$, from the third of Eqs. (\ref{eq2:13}) and from
Eq. (\ref{eq2:49}) we obtain the matrix

\begin{equation}
D^{\prime}(a_0g)=D^{\prime}(a_0)D^{\prime\ast}(g)=
\left(\begin{array}{r|r}
0 & N\Delta^{\ast}(g)
\\
\hline
-N\Delta^{\ast}(g) & 0
\end{array}\right)
\label{eq2:51}
\end{equation}

\noindent
When the similarity transformation in Eq. (\ref{eq2:18}) is applied to the above matrices
connected with the coset $a_0G$, they acquire the factor ${\rm exp} (i\alpha_0)$.
\\
\indent
We observe that Eqs. (\ref{eq2:28}), (\ref{eq2:50}) and (\ref{eq2:51}) can be obtained by 
applying to the corep matrices in Eqs. (\ref{eq2:11}) and (\ref{eq2:12}) the transformation:

\begin{equation}
V_2=\left(\begin{array}{r|r}
iE & 0
\\
\hline
0 & iN^{-1}
\end{array}\right)
\label{eq2:52}
\end{equation}

\noindent
When the matrix $N$ is unitary, $V_2$ turns into the transformation given
by Kovalev and Gorbanyuk in Eq. (8.10) of \cite{Kovalev6}, (or
Eq. (1.5.50) of \cite{Kocinski3}).
\\
\indent
The basis functions transforming according to the matrices in Eqs. (\ref{eq2:28}), 
(\ref{eq2:50}) and (\ref{eq2:51}) are determined from the equality,

\begin{equation}
\tilde V_2\left(\begin{array}{r}
\varphi
\\
\phi
\end{array}\right)=\left(\begin{array}{r}
i\varphi
\\
i{\tilde N}^{-1}\phi
\end{array}\right)=\left(\begin{array}{r}
\psi_1
\\
\vdots
\\
\psi_{2d}
\end{array}\right)
\label{eq2:53}
\end{equation}

\noindent
with $\phi_i$ in (\ref{eq2:46}) or, 
\begin{eqnarray}
\psi_j=i\varphi_j,\quad j=1,...,d
\nonumber\\
\psi_{d+j}=-i\sum\limits_{k=1}^d (\tilde N^{-1})_{jk}\varphi_{d+k},\quad j=1,...,d,\quad
\mbox{with}\quad \varphi_{d+k}=a_0\varphi_k
\label{eq2:54}
\end{eqnarray}

\noindent
The corepresentation formulas hold for single-valued as well as for
double-valued representations $\Gamma$ of the subgroup $G$.

\section{Conclusions}

We have presented the corepresentation theory without the assumption
of the unitarity of the subgroup $G$ of the group $G+a_0 G$,
where $a_0$ denotes an antilinear operation.
The formulas of the corepresentation theory with unitary groups $G$ can be obtained from this presentation. 

\section*{Acknowledgments}
We are indebted to Professor Zbigniew Oziewicz from the Universidad
Nacional Aut\'onoma de M\'exico for the discussions concerning the mappings with antilinear operations.

\end{document}